\documentclass[prd,nofootinbib,twocolumn]{revtex4-1}
\usepackage{graphicx}
\usepackage{color}

\usepackage{hyperref}

\newcommand{\code}[1]{\texttt{#1}}

\begin{document}

\title{The Current Status of Binary Black Hole Simulations in Numerical
  Relativity}

\author{Ian Hinder}
\affiliation{
Max-Planck-Institut f\"ur Gravitationsphysik,
Albert-Einstein-Institut, \\ 
Am M\"uhlenberg 1, D-14476 Golm, Germany}

\date{\today}

\begin{abstract}
  Since the breakthroughs in 2005 which have led to long term stable
  solutions of the binary black hole problem in numerical relativity,
  much progress has been made.  I present here a short summary of the
  state of the field, including the capabilities of numerical
  relativity codes, recent physical results obtained from simulations,
  and improvements to the methods used to evolve and analyse binary black
  hole spacetimes.
\end{abstract}

\maketitle

\section{Introduction}

With the current generation of gravitational wave interferometers
(LIGO \cite{Abbott:2007kv, Waldmann:2006bm, ligoweb}, Virgo
\cite{Acernese:2006bj, virgoweb}, GEO \cite{Danzmann:1992, Hild:2006bk,
  geoweb} and TAMA \cite{Takahashi:2004rwa}) reaching their design
sensitivity, and the next generation (Advanced LIGO
\cite{Fritschel:2003qw} and LISA \cite{Danzmann:1998,
  Danzmann:2003tv}) planned to come online in the next few years, the
accurate modelling of
gravitational wave sources and the construction of template waveforms
to be used in matched filtering of detector data has become urgent.
One of the strongest expected sources of gravitational waves is the
inspiral and merger of a binary black hole (BBH) system.  Additionally, the
modelling of BBH mergers has applications in astrophysics, where it
can be used to learn about possible growth methods for supermassive
black holes,
galaxy evolution and globular clusters.
BBH mergers occur in the regime of very strong gravity, and whilst the
early inspiral can be
described accurately with approximate analytic post-Newtonian (PN)
methods, to model the late inspiral and merger requires the numerical
solution of the Einstein equations using expensive and complex
supercomputer simulations in numerical relativity (NR).  Though the analytic framework and
associated numerical simulations have been in development since the
1960s, it was only in 2005 that simulations of multiple orbits of a
BBH system became possible.
Since those first simulations
\cite{Pretorius:2005gq,Baker:2005vv,Campanelli:2005dd}, the quality of
the computations has improved significantly,
and it is now possible to simulate enough orbits with a sufficiently
high accuracy that the resulting waveforms can be compared with
PN results.  NR waveforms can
be attached to PN waveforms to construct {\em hybrid} waveforms, fitted
with {\em phenomenological} models, or used to tune
Effective-One-Body (EOB) waveforms.

In this brief status report, I review recent developments in the
field of numerical relativity.  As such, this work is not intended to
be a comprehensive discussion of the history of the subject.  For more
details on earlier work, particularly associated with waveforms for
gravitational wave (GW)
data analysis, see Ref.~\cite{Hannam:2009rd}.  Several
works have appeared as this manuscript was being prepared, and this demonstrates
the healthy level of activity in the field.
This article is structured as follows.
In Sec.~\ref{sec:stateoftheart}, I give a brief overview of the
current capabilities of NR codes, including the longest waveforms and
highest mass ratios studied so far.  I also summarise the results of a
validation study (the Samurai project) which verified the consistency
of the waveforms produced by the several codes in the community, and
conclude the section with a discussion of some work on improvements to
computational efficiency.

In Sec.~\ref{sec:physics}, I discuss new physics results which have
been obtained using NR simulations, including comparisons with PN and
EOB models.  I then discuss some applications of
NR to gravitational wave data analysis, including work on injecting
numerical waveforms into data analysis pipelines (the NINJA project),
the construction of phenomenological waveform templates, and studies of
NR results applied to the
detection of gravitational waves and the estimation of the parameters
of their sources. I then discuss work
that has been done on modelling the state of the final merged black hole
as a function of the parameters of the initial black holes, as
well as work on hyperbolic (non-circular) and high velocity BBH
configurations.

In Sec.~\ref{sec:methods}, I present recent work on methods used in NR
for BBH simulations.  Proposed new techniques for generating improved
initial data are reviewed, as well as work which has been done on
boundary conditions.  I then
discuss studies involving the computation of gravitational waves from
numerical
simulations, including
analyses of finite radius effects and extrapolation procedures, the
asymptotic falloff of the Weyl scalars used for measuring
spacetime content including gravitational radiation, and the first
application of Cauchy-Characteristic Extraction (CCE) to produce
unambiguous waveforms at future null infinity for BBH systems.  I also
discuss the development of new computational infrastructure for using
non-Cartesian grids with finite differencing BBH codes, as well as
improvements to the methods used for spectral BBH evolutions.

In Sec.\ref{sec:challengesandconclusions}, I list the main challenges
facing the NR community today, and present some closing remarks.

\section{State of the Art}
\label{sec:stateoftheart}


There are now several groups around the world with codes capable of
performing BBH simulations \cite{Pretorius:2005gq,
  Imbiriba:2004tp, Campanelli:2005dd, Vaishnav:2007nm,
  Sperhake:2006cy, Brugmann:2008zz, Scheel:2006gg, Pollney:2007ss,
  Faber:2007dv, Yamamoto:2008js, Cao:2008wn, Liebling:2002qp}. 
There are two analytic forms of the Einstein
equations (BSSN \cite{Nakamura:1987zz, Shibata:1995we,
  Baumgarte:1998te} and Generalised Harmonic
\cite{Pretorius:2004jg, Pretorius:2005gq, Lindblom:2005qh}), and various
numerical methods (high order finite differencing, pseudo-spectral,
adaptive mesh refinement, multi-block) in use in the community.

\subsection{Consistency}

The
gravitational waveforms computed by these codes for a given set of
initial data should be independent of these differences as they represent
a physical observable, so a comparison of the results of the different
codes yields a strong test of consistency.  The Samurai
\cite{Hannam:2009hh} project was a collaborative effort to compare the
waveforms from five different codes for
the last 4 orbits and merger of a binary of equal-mass, non-spinning
black holes in circular orbit.  The codes were \code{BAM}
\cite{Brugmann:2008zz}, \code{CCATIE} \cite{Pollney:2007ss}, \code{Hahndol}
\cite{Imbiriba:2004tp,vanMeter:2006vi}, \code{MayaKranc}
\cite{Vaishnav:2007nm} and \code{SpEC} \cite{Scheel:2006gg}.  The first four
use the BSSN 
formulation of the Einstein equations, though the coordinate
conditions used are slightly different for each code.  \code{SpEC} uses the
Generalised Harmonic formulation.  \code{CCATIE} and \code{MayaKranc} both use the
same computational framework (\code{Cactus}/\code{Carpet}
\cite{Goodale02a,Cactusweb,Schnetter:2003rb,carpetweb}),
but the
evolution and analysis codes are separate.  The other codes used in
the Samurai comparison are
independent.  \code{SpEC} uses a pseudo-spectral evolution
scheme, whereas the others use finite differencing methods.  Despite
these differences, it was found that the waveforms all agreed
with each other within their published error estimates.  In
gravitational wave data analysis, the quantity usually used to make
comparisons between waveforms is the {\em match}, $\mathcal{M}$ \cite{Owen:1995tm}.  This
is a normalised frequency-domain scalar product of the waveforms,
weighted by the detector noise, in the range zero to one, where a
match of $\mathcal{M} = 1$ indicates two waveforms which are seen as
the same by a
given detector.  We also refer to the {\em mismatch}
$1 - \mathcal{M}$ of two waveforms.  The mismatch between the Samurai
waveforms was less than $10^{-3}$ for enhanced LIGO, Advanced LIGO,
Virgo and Advanced Virgo, suggesting that the accuracy and consistency
of the waveforms is sufficient for detection purposes.  We should note
here that, since only the last eight cycles of the waveform were
considered, this result applies only to those systems where these
cycles are the only ones in the detector bandwidth; i.e.~those systems
above a certain total mass.  Also, only the dominant $l = 2, m = 2$ multipolar
mode was considered, approximately corresponding to a system observed
from a direction perpendicular to the orbital plane.  The impact of
the inclusion of higher modes and longer waveforms (e.g.~constructed
by joining NR to PN waveforms) in such a comparison has not yet been
studied.


\subsection{Longest Waveform} The longest NR BBH waveform so far produced lasts for 15 orbits and
includes the merger and ringdown phases, and is described in Boyle et
al.~\cite{Boyle:2007ft} and Scheel et al.~\cite{Scheel:2008rj}.  This waveform, from an equal mass binary of
non-spinning black holes, was generated using the \code{SpEC} code and, due to its length and quoted
accuracy, has been used in a number of studies comparing NR and PN
results \cite{Damour:2009kr, Buonanno:2009qa,
  Damour:2007yf,Boyle:2008ge, Boyle:2009dg, Pan:2009wj}.  This code uses a pseudo-spectral numerical method and
multiple coordinate patches to cover the evolution volume and, in
Szil{\'a}gyi et al.~\cite{Szilagyi:2009qz}, new techniques were introduced which allow
simulation of the full inspiral, merger and ringdown without the
fine-tuning of parameters that was previously required.  As a result
of this work, simulations of BBH mergers with mass ratios
$q = m_1 / m_2 \ge 1$ (where $m_1 \ge m_2$ are the
masses of the individual holes) of $q = 2$ and
dimensionless spins up to 0.4 are now possible with the \code{SpEC} code, and
these are presented in Ref.~\cite{Szilagyi:2009qz}.  Additionally,
simulations with dimensionless spins of 0.44 anti-aligned with the
orbital angular momentum are presented in Chu et al.~\cite{Chu:2009md}.  
In a talk by H.~Pfeiffer \cite{tlk:pfeiffernrda2009}, a series of long unequal
mass simulations performed with the \code{SpEC} code was presented, with 15
orbits up to $q = 4$ and 8 orbits up to $q = 6$. 


\subsection{Highest Mass Ratio}

For a 3D numerical relativity code, one of the most challenging
problems is the simulation of black holes of very different
masses. Unequal masses are expected to be commonplace astrophysically, 
hence waveforms from these systems are essential.
One early reason for the interest in unequal mass BBH systems, besides the
use of NR waveforms for GW science, was the fact that the
mass asymmetry leads to an asymmetry in the linear momentum emitted in
the gravitational waves around the merger, leading the final black hole to recoil
out of the initial zero momentum frame.  This recoil
(also known as a {\em kick} or {\em rocket effect}),
especially when the black holes are spinning, can have important consequences
for astrophysics \cite{Merritt:2004xa}.  
The highest mass ratio that has been simulated is $q = 10$.
The first such simulation, reported in Gonz\'ales et al. \cite{Gonzalez:2008bi}, consisted of three orbits of a
binary of non-spinning black holes as well as the merger and ringdown of the
final black hole to Kerr.  The result was computationally very expensive to
achieve, but it verified the prediction of the empirical recoil
formula \cite{Gonzalez:2006md, Baker:2008md, Schnittman:2007sn} in a range
which had previously not been tested.  The reason for the increase in
computational expense is that, for a fixed total mass $M = m_1 + m_2$,
the gravitational wavelength remains approximately constant with
varying $q$, but the length and time scale required to resolve the
smaller hole scales approximately with $q$.   With adaptive mesh
refinement as used in Ref.~\cite{Gonzalez:2008bi}, for large $q$, the computational
cost in CPU hours is $q$ times higher than for an equal mass
simulation.  One technical
problem which arises when simulating unequal mass BBH systems relating
to the coordinate conditions used in many codes
has been studied recently in M\"uller et al.~%
\cite{Mueller:2009jx}.  Due to the computational expense of simulating
high mass ratios, it can be desirable to propagate the gravitational
waves in a separate evolution.  In Lousto et al.~\cite{Lousto:2010tb}, a BBH
simulation with a mass ratio of 1:10 was presented, along with a
computationally inexpensive
perturbative evolution of the gravitational waves modelled using point
particles as the sources, whose locations were determined from the
coordinate tracks of the black holes in the numerical simulation.  


\subsection{Highest spin}
Mathematically, Kerr black holes have a maximum dimensionless spin of
$1$, and there is a good probability that highly spinning black holes exist in
nature \cite{Volonteri:2004cf, Gammie:2003qi, Shapiro:2004ud}.
Current NR techniques for simulating BBH systems are
limited in the maximum spin which can be specified in the initial
data, both for theoretical and numerical reasons.  The common initial
data formulations have mathematical limits on the spins which can be
achieved (see Sec.~\ref{sec:idspins} for a discussion of techniques
for constructing initial data with high spins).  Numerically, black holes
with high spins require more resolution and hence higher computational
resources to achieve a given accuracy.  Typically, black holes with
dimensionless spins as high as 0.6--0.8 can be evolved with only a
moderate increase in computational cost over the non-spinning case.
The cost of simulating higher spins increases dramatically however.
For the Bowen-York initial data used by the majority of BBH codes, the
theoretical maximum has almost been reached in Dain et al.~\cite{Dain:2008ck},
where $\sim 7.5$ orbits of black holes with dimensionless spins $0.92$ are
evolved.  This is the highest spin for the inspiral-merger-ringdown of
a BBH system so far simulated in NR.

\subsection{Lowest Eccentricity} 
BBH systems visible to GW detectors are usually
expected to have circularised to an orbit of eccentricity zero due to
the emission of gravitational radiation \cite{Peters:1964zz}.
However, it is not known how to rigorously choose BBH initial data
parameters so that the resulting orbit is quasi-circular.
The lowest eccentricities have been obtained using a method presented in
Pfeiffer et al.~\cite{Pfeiffer:2007yz}.  In this method, the first few orbits of a BBH
evolution are performed and the initial data parameters adjusted in an
iterative procedure based on the measured eccentricity of the
resulting evolution. With this method, eccentricities as low as $e
\sim 5 \times 10^{-5}$ have been produced.  This procedure must be
repeated for every initial data set considered, though the
computational cost is mitigated by the fact that the evolutions do not
need a very high resolution.
It should be noted that there is no fully general relativistic
definition of eccentricity; definitions can be used which are based on
the coordinate or proper separation of the black hole horizons, and
these definitions are used in Ref.~\cite{Pfeiffer:2007yz}.  It would
also be possible, though more difficult, to use the properties of the
gravitational radiation, such as its amplitude or frequency.


\subsection{Computational Performance}

As higher mass ratios and longer simulations are required, the
performance of NR codes becomes more important.  The XiRel project
\cite{ES-Tao2008a,xirelweb} was started in order to improve the performance of the publicly available
Carpet \cite{Schnetter:2003rb,carpetweb} adaptive mesh refinement infrastructure.  As a result of
recent work, Carpet now scales efficiently up to 2048 processing
cores, and as it is used for BBH simulations at AEI, GaTech, LSU, RIT
and UIUC, the improvements benefit simulations by all of these groups.

\section{New Binary Black Hole Physics}
\label{sec:physics}

\subsection{Comparisons with Analytic Models}

Due to the high computational cost of NR BBH simulations, the early
inspiral can instead be modelled using approximate analytic techniques based on the
post-Newtonian method (see Sch\"afer \cite{Schaefer:2009dq} for a recent
review), and the late inspiral and merger can be simulated with NR.
PN results are accurate in the regime where the black holes are far apart and
moving slowly.  For GW detection purposes, the use of longer waveforms increases
the range of masses of systems which can be detected, so
including the inspiral is very important.  Recent work has focused on
combining the waveforms from analytic methods with NR simulations to
generate complete waveforms describing both the inspiral and merger.
For a review of the first PN-NR comparisons see, for example,
Hannam \cite{Hannam:2009rd}.

The Effective-One-Body (EOB) approach \cite{Buonanno:1998gg} (see
Damour and Nagar \cite{Damour:2009ic} for a recent review) is heavily based on the PN
method, but uses additional techniques to model the merger phase,
which is usually inaccessible to PN methods.  There are several
unknown parameters in the various EOB methods, and these need to be
determined by comparison with NR simulations.  Two recent works,
Damour et al.~\cite{Damour:2009kr} and Buonanno et al.~\cite{Buonanno:2009qa}, have used the \code{SpEC}
15 orbit equal mass non-spinning waveform \cite{Boyle:2007ft, Scheel:2008rj} to tune
the parameters in the EOB formalism, adding to previous work on PN and
EOB comparisons with this waveform in
Refs.~\cite{Damour:2007yf,Boyle:2008ge}.
After this tuning, the EOB model
agreed with the NR waveform within the quoted numerical error on the
NR waveform.  For binaries with total masses between 30 and 150
$M_\odot$
(solar masses), there was a mismatch between NR and EOB of $< 10^{-3}$ for LIGO, Enhanced LIGO and Advanced LIGO.
This work was very recently extended in Pan et al.~\cite{Pan:2009wj} to include
the new \code{SpEC} waveform \cite{Chu:2009md} incorporating spinning
black holes.

In Campanelli et al.~\cite{Campanelli:2008nk}, a 9-orbit NR simulation of a configuration of
spinning black holes of unequal mass with their spin axes misaligned
with the orbital angular momentum was presented.  This is the most
general BBH configuration that can be expected in nature, assuming
that such systems will have evolved to have zero eccentricity by the
time the NR portion of the waveform enters the frequency range of GW
detectors.  The mass ratio was $q = 1.25$ and the dimensionless spins
of the black holes were 0.6 and 0.4.  In this simulation, significant
precession of the orbit out of the initial orbital plane was observed,
in agreement with PN spin-orbit and spin-spin coupling predictions (see,
e.g.~Kidder \cite{Kidder:1995zr} for details).
A comparison with 3.5 PN waveforms including spin
effects was made, and the first 6 wave cycles (3 orbits) agreed within
1\% (note that this is a time-domain inner-product measure, not the
conventional frequency-domain match $\mathcal{M}$ usually used in
gravitational wave data analysis).

The work of Peters \cite{Peters:1964zz} in the 1960s demonstrated that
a binary system in an eccentric orbit which is inspiralling due to the
emission of gravitational radiation will lose eccentricity and
eventually circularise.  Typical gravitational wave sources are
expected to circularise before they can be observed, so studies of
sources of gravitational radiation usually assume circular orbits.
However, it has been
suggested \cite{Wen:2002km, Benacquista:2002kf, Gultekin:2004pm, Blaes:2002cs, Dotti:2005kq} that there may be sources with non-negligible
eccentricity which are detectable by Advanced LIGO and LISA.  
In
Hinder et al.~\cite{Hinder:2008kv}, the first comparison between an eccentric PN model
and an NR simulation was performed.  A 10 orbit simulation (20 GW
cycles) with initial eccentricity $e \approx 0.1$ was presented, and
an eccentric PN model was fitted to the resulting waveform.
The two waveforms agreed within 0.1 radians for the first 8 cycles,
and by 5 cycles before merger, a dephasing of 0.8 radians had occurred.
Two different PN models were compared, and the
choice of PN expansion variable was found to have a significant effect
on the accuracy of the PN approximation.

\subsection{Gravitational Wave Data Analysis}

Due to the weak nature of gravitational waves, and the difficulties in
separating GW signals from local vibrations, the output of
ground-based gravitational wave
detectors is dominated by noise.  As a result, sophisticated
statistical algorithms must be used to extract physical signals
corresponding to the detection of gravitational waves from BBH
systems.  These algorithms, employing the technique of {\em matched
  filtering}, require accurate waveform {\em templates} corresponding
to the sources to be detected.


For low mass systems (those with a total mass of between 2 and 35
$M_\odot$), current GW detector data is searched for signals using PN
template waveforms.  This is because for high masses, the GW frequency
of the merger is too high to be seen by the detectors, whereas for low
masses, pure PN templates can be sufficient \cite{Buonanno:2009zt}.
Between 25 and 100 $M_\odot$, waveforms generated using the EOB
formalism (not including spin effects) and tuned to NR (EOBNR
\cite{Buonanno:1998gg, Buonanno:2000ef, Damour:1997ub, Damour:2000we})
as well as phenomenological templates \cite{Ajith:2007kx,
  Ajith:2007qp} are currently used.  For a recent status report on
the searches of LIGO-Virgo data for signals from coalescing binaries,
see Sengupta et al.~\cite{Sengupta:2009nm}, and for more information on the use of NR
waveforms in GW data analysis, see Hannam \cite{Hannam:2009rd}.

Now that NR waveforms are available which include the
merger phase, as well as a significant overlap with the regime of PN
validity, it is important to test that the search algorithms work well
with signals which include the merger.  This work was started in
the Numerical Injection
Analysis (NINJA) project \cite{Aylott:2009tn, Aylott:2009ya, ninjaweb}.  NR waveforms from ten
different groups were injected into the data analysis pipeline
software and the performance for detection and parameter estimation of
these pipelines was characterised.  
The NR waveforms covered mass ratios up to $q = 4$, a range of spin
configurations up to dimensionless spins of $\sim 0.9$, and up to 30
GW cycles.  The waveforms were converted into time-series data that
would be seen at the detector and Gaussian noise, designed to mimic
the features of each detector, LIGO, Virgo and Geo600, was added to
the signal.  Various data analysis techniques were then used to
attempt to detect the signals in the data, and in some cases to
estimate the source parameters.  Overall, it was found that many of
the current data analysis pipelines were able to detect the merger
waveforms at the expected sensitivities, but that significant work is
needed to improve parameter estimation.  Additionally,
the NINJA project has provided a working model for collaboration and
communication between the NR and DA communities; essential if the
recent advances in NR are to contribute to GW science.  The NINJA 2
project \cite{ninjaweb}, which is currently underway, will incorporate
hybrid
waveforms (i.e.~PN and NR waveforms combined), and real detector
noise.  This will allow stronger statements to be made concerning the
performance of the detector pipelines for real signals.


In Reisswig et al.~\cite{Reisswig:2009vc}, a study was made of the signal to
noise ratio (SNR) of a family of waveforms for BBH systems with spins
aligned and anti-aligned with the orbital angular momentum.  In this
family, there is no precession of the orbital plane.  It was found
that the SNR increases with the projection of the total spin in the
direction of the orbital angular momentum, and that if the spins are
aligned with the orbital angular momentum and maximal (dimensionless
spin of 1),
the SNR is three times as high as in the anti-aligned case, leading to an
increase in the event rate for current and advanced LIGO and Virgo
detectors of a factor of $\sim 30$.  Hence it is likely that there
will be a bias in detected binaries towards high spins aligned with
the orbital angular momentum, as these have the largest SNRs.  It was
also determined that the match between different cases is controlled
by the total spin; binaries with zero total spin (including the case
with both black holes non-spinning) are indistinguishable from each other
(indicated previously in Vaishnav et al.~\cite{Vaishnav:2007nm}) for LIGO, whereas
binaries with a nonzero total spin have identifiably different
gravitational wave signatures.  This means that aligned spin detection
templates only need a single spin parameter, which reduces the
dimensionality of the search parameter space.


In Ajith et al.~\cite{Ajith:2009bn}, it is shown that if only
non-spinning templates are used to search for signals from spinning
systems, the event rate is reduced by up to 50\%.  To alleviate this
problem, it is necessary to include spin effects in the templates used
in GW data searches.  The authors constructed time-domain hybrid
waveforms by matching waveforms from NR simulations covering at least
8 waveform cycles to a PN approximant called {\em TaylorT1}
\cite{Damour:2000zb}.
Mass ratios up to $q =
3$ and dimensionless spins up to $0.85$ aligned and
anti-aligned with the orbital angular momentum were considered.
Phenomenological Fourier-domain waveform models with free parameters
were then constructed, and the parameters were fitted to the PN-NR
hybrids.  This resulted in a bank of template waveforms with freely
adjustable mass ratio and spin parameters which can be used in a GW
data-analysis pipeline, and should lead to higher detection rates for
spinning BBH systems.


Once a detection has been made, the next step is to determine the
parameters of the source, for example the sky location, mass ratio,
total mass, etc.  For the space-based detector LISA, the low noise
levels mean that detection of BBH signals is generally not a problem; the
challenge is to accurately determine the source parameters.  In
McWilliams et al.~\cite{McWilliams:2009bg}, a comparison is made between the accuracy of parameter
estimation when using purely PN waveform models and when using
combined PN and NR models.  It is found that the inclusion of the NR
merger portion for supermassive BBH systems (for mass ratios $q < 10$)
reduces the uncertainty in the sky-location of such a binary by a
factor of $\sim 3$ to within $\sim 10$ arcminutes.

\subsection{Modelling the Final State}

Knowledge of the mass, spin and linear momentum of the final black hole in a
BBH merger has many applications.  One example is in the modelling of
the evolution of cosmological black hole spins \cite{Berti:2008af}.  Another
is in explaining the presence or absence of massive black holes in the centre
of galaxies, as the merger of galaxies and their associated central
massive black holes may result in recoils which eject the final black hole
\cite{Merritt:2004xa}.

When it was predicted and then confirmed
\cite{Campanelli:2007ew,Gonzalez:2007hi,Campanelli:2007cga} that very
large recoils could be generated in BBH mergers with specific spin
orientations, a large amount of effort went into exploring this effect
due to the potential application to astrophysics.  Recently, in
Lousto et al.~\cite{Lousto:2009mf}, the mass, spin, and recoil velocity of the final
merged black hole were modelled as functions of the spins just before the
merger using expressions based on PN predictions, but with PN coefficients
replaced by those obtained from fits to NR data.
This is an extension of previous
work \cite{Campanelli:2007ew,Baker:2008md,Lousto:2008dn} including
additional terms in the models.
In Refs.~\cite{Boyle:2007sz, Boyle:2007ru}, the authors proposed
formulae for properties of the final black hole (mass, spin, linear
momentum) based on symmetry arguments and an expansion in the initial
spin parameter.  The final spin has more recently been addressed using
an alternative approach (also see Refs.~\cite{Hughes:2002ei,
  Buonanno:2007sv, Kesden:2008ga} for predictions based on
point-particles).  
In Barausse and Rezzolla \cite{Barausse:2009uz}, following work in
Refs.~\cite{Rezzolla:2007xa, Rezzolla:2007rd, Tichy:2008du,
Rezzolla:2008sd},
a set of approximations and
simplifying assumptions is combined with a large body of
already-published NR data to produce a general formula for the
spin magnitude and direction of the final black hole after a merger.  Instead
of basing the model on PN expressions around the time of the merger,
the final spin is expressed in terms of the spins early in the
inspiral, and it is argued that this is more relevant when making
astrophysical predictions, since an astrophysicist is not interested
in the detailed dynamics of the merger, but instead in the final state
obtained from a set of initial parameters when the black holes are
far-separated.  In Lousto et al.~\cite{Lousto:2009ka}, PN evolutions and the 
empirical model of Ref.~\cite{Lousto:2008dn} are combined to make a link in a statistical sense
between the early inspiral and pre-merger spins.


While the above works have concentrated on determining the final state
of the black hole using empirical models, in Lovelace et al.~\cite{Lovelace:2009dg}, the
dynamical momentum flow in the region near the black holes is studied.  The
method used is based on field theory on an auxiliary spacetime in
which linear momentum is defined, and follows on from previous work
which treated the PN case \cite{Keppel:2009tc}.  In this way, the
linear momentum of
the individual black holes can be computed, albeit with some gauge dependence.
It is found that, despite this gauge dependence, the linear momentum
measured for a head-on collision of two black holes using this method
agrees well between evolutions using different gauges, suggesting that this linear
momentum measurement may have a physical interpretation.

\subsection{Beyond Circular Inspirals}

Due to its astrophysical importance, the BBH problem in NR is usually
studied for quasi-circular configurations; i.e.~those consisting of
circular motion combined with a low inward radial velocity. However,
there have recently been several works which study non-quasi-circular
configurations.  


Hyperbolic BBH encounters can be thought of as orbits of eccentricity
$e > 1$. In a Newtonian system, such a configuration would result in
scattering of one black hole off the potential of the other, but in full GR,
for a sufficiently small impact parameter, the black holes become
gravitationally bound due to the emission of energy through
gravitational waves and merge quickly \cite{Pretorius:2007jn}.
In Healy et al.~\cite{Healy:2008js}, it was found that hyperbolic encounters of
spinning black holes resulted in a significant linear momentum imparted to the
final black hole due to the asymmetry of the emitted radiation at the moment
of the merger.  Recoil velocities as high as 10000 km/s could be
obtained, in contrast with the quasi-circular case where the highest
recoil velocity obtained is of the order of 3300 km/s.  In
Healy et al.~\cite{Healy:2009ir}, it was found that the final
dimensionless spin
could be as high as 0.98 when extrapolated to the case of
maximally spinning black holes.


In particle physics, high energy collisions are used to probe the
fundamental properties of nature.  As these collisions are performed
at higher and higher energies, it is expected that they will result in
the formation of an event horizon, and that these scattering processes
will be well-modelled by the high velocity scattering of black
holes. In Sperhake et al.~\cite{Sperhake:2008ga}, the authors studied high velocity
head-on black hole collisions.  The simulations included initial velocities
up to $v = 0.94 c$ and, when extrapolated to black holes moving at the speed of
light, the energy radiated in the head-on collision was determined to
be about 14\%.  In Shibata et al.~\cite{Shibata:2008rq}, the authors studied
non-head-on high velocity collisions.  By varying the angle of
approach of the black holes, it was determined that the impact parameter
should be approximately $b < 2.5 G M/c^2/v$ for a black hole to form,
where $M$ is the total black hole mass.  This model was
fitted from the simulated velocities from $0.6 c$ to $0.9 c$.  A
similar study was performed in Sperhake et al.~\cite{Sperhake:2009jz} where the result
was confirmed for these velocities, but for higher velocity ($v =
0.94$), the formula in Ref.~\cite{Shibata:2008rq} may have been an overestimate.


Zoom-whirl orbits are characterised by quasi-circular motion (whirl),
alternating with highly eccentric motion (zoom).  For BBH systems,
this phenomenon has
been seen in extreme mass ratio inspirals as well as PN models, and in
Pretorius and Khurana \cite{Pretorius:2007jn} it was seen for the first time in NR. In
Sperhake et al.~\cite{Sperhake:2009jz}, zoom-whirl behaviour is confirmed for the high
velocity collisions studied, and in Healy et al.~\cite{Healy:2009zm}, the authors
study zoom-whirl orbits for a variety of mass ratios and spins and
show that they can occur without the need to fine-tune the initial
data parameters.  Furthermore, in Gold and Br\"ugmann \cite{Gold:2009hr}, low momentum zoom-whirl
orbits, expected to more closely reflect astrophysical scenarios, are
simulated and the dependence of the energy radiated as a function of
the initial angle of the black hole momenta is computed.  
They find that as the initial angle between the momenta and the
coordinate line joining the two BHs increases from zero, there is a
peak in the emitted energy at 47 degrees, followed by several local
maxima.

\subsection{Electromagnetic Counterparts}

Several recent works have studied possible electromagnetic (EM) signatures
of BBH mergers.  If such signatures are
visible to current or future instruments, this opens up the
possibility of coincident detection via EM and GW observations of BBH
inspiral events.  Not only might this increase the detectability of
these events, but it might significantly increase the accuracy with
which the parameters of the binary, such as the sky location and redshift, can be
determined.

In Palenzuela et al.~\cite{Palenzuela:2009yr, Palenzuela:2009hx}, the fully coupled
Einstein-Maxwell system is evolved for an equal mass, non-spinning BBH
system in the presence of an external magnetic field but otherwise in vacuum.  The system is
designed to model the effect of the magnetic field sourced by an
accretion disk around a supermassive black hole binary, where the matter from
the accretion disk has been cleared from the immediate neighbourhood
of the black holes by binary torques \cite{Milosavljevic:2004cg}.
It is found that the EM fields evolve in a way which follows the
dynamics of the binary and so they can be used as {\em tracers} of the
motion.  Additionally, the EM energy flux is found to oscillate with a
quarter the orbital period of the binary and the energy in the EM
field is found to increase as the binary approaches merger.  For the
astrophysically realistic field strengths used here, the addition of
the EM field has no effect on the binary dynamics or waveform.  In
followup work by M\"osta et al.~\cite{Mosta:2009rr}, therefore, the EM fields are
treated as test fields and their evolution does not feed back into the
evolution of the geometry.  Here, larger initial black hole separations are
used along with black hole spins aligned and anti-aligned with the orbital
angular momentum.  
It is found that
for the lowest multipolar modes, the EM radiation follows very closely
the amplitude and phase evolution of the GW radiation, so that
detecting either one gives direct information about the
other. Furthermore, the EM energy flux scales
quadratically with the black hole spin, and is 13 orders of magnitude lower
than the GW energy flux. Most importantly, the frequencies of the EM
radiation are well outside the range of existing radio observations
and hence direct observation is highly unlikely.

In an alternative picture of the late inspiral of a supermassive black hole
system, if the surrounding gaseous environment is sufficiently hot, as for example in the
nuclear regions of some low luminosity active galactic nuclei (AGNs),
there could be gas present near the black holes all the way up to and through
merger \cite{Elitzur:2009db}.
In van Meter et al.~\cite{vanMeter:2009gu}, first steps are taken
towards adding a gas into their BBH simulations by computing geodesics
of the spacetime, and considering these to represent flows of test
particles on the background geometry of the BBH solution.  Differences
in the collision and outflow speeds of the test particles are observed
between single black hole and BBH systems, and between spinning and
non-spinning black holes in binaries.
In Bode et al.~\cite{Bode:2009mt} and, independently, in Farris et al.~\cite{Farris:2009mt},
a supermassive BBH system is simulated in the
presence of a gas modelled using full GR hydrodynamics.  In Ref.~\cite{Farris:2009mt}, the rate of accretion of
the gas onto the black hole is computed, and it is found that the luminosity
of the resulting EM radiation is enhanced significantly over that
expected from a single black hole, and estimates of detectability of this
radiation are given.  In Ref.~\cite{Bode:2009mt}, correlations are found
between the EM and GW emission.  For the most massive supermassive black hole
systems that would be detectable by LISA, the EM luminosities are here
found to be high enough to be accessible to observations out to a
redshift of $z \approx 1$ using X-ray measurements.

\section{New Binary Black Hole Methodology}
\label{sec:methods}

\subsection{Initial Data}
\label{sec:idspins}

Specification of initial data for a BBH evolution is a nontrivial task
as this data must satisfy the Einstein constraint equations.  These
equations are nonlinear so it is not in general possible to
construct initial data by superposing two single black hole solutions.  There
are two main methods for constructing BBH initial data in use for NR
evolutions; Bowen-York puncture data \cite{Bowen:1980yu,Brandt:1997tf}
and a method based on the Conformal Thin
Sandwich (CTS) decomposition of the equations involving excision of
the black hole interiors from the domain \cite{York:1998hy, Pfeiffer:2002iy,
  Cook:2004kt, Caudill:2006hw, Pfeiffer:2007yz}.  In both these
cases, very roughly speaking, one chooses the desired characteristics of the black holes and
then solves a set of elliptic equations numerically.  Both of these
methods are subject to several problems: incorrect initial radiation
content, difficulty in controlling eccentricity, and difficulty in
including very high spins.

{\em Radiation content:} In the majority of current simulations, the
initial data contains spurious radiation (also known as {\em junk}
radiation) in the
neighbourhood of the black holes, due to simplifications made when
constructing the initial data (in particular the conformal-flatness
assumption).  It also does not contain the radiation
which would have resulted from the evolution of the binary from early
times.  The spurious radiation is visible in BBH waveforms only at early
times, and is followed by the astrophysically expected waveform.
Typically the initial part of the waveform must be discarded, and the
spurious radiation can also cause unphysical numerical reflections from
mesh refinement or outer boundaries.  In 
Johnson-McDaniel et al.~\cite{JohnsonMcDaniel:2009dq}, the authors propose a new method of
constructing initial data based on the PN approximation.  This initial
data set is not conformally flat, and contains the radiation expected
from the previous inspiral.  One drawback of this method is that the
Einstein constraint equations are satisfied only to a given PN order.
In Kelly et al.~\cite{Kelly:2009js}, another type of initial data, also constructed
using the PN method \cite{Kelly:2007uc}, is evolved.  This initial
data is also constructed to satisfy the constraint equations to a
given PN order, as with previous approaches
\cite{Nissanke:2005kp}.
The resulting
evolution gives the expected waveform from the start and contains
significantly less spurious radiation than the evolution of comparable
Bowen-York initial data.  On the other hand, there is significantly
higher orbital eccentricity ($e \sim 0.1$), as well as a variation in
apparent horizon mass, perhaps caused by the constraint-violating
nature of the initial data.

{\em Eccentricity:} Recently, in Walther et al.~\cite{Walther:2009ng}, a PN method was used for
constructing low eccentricity initial data parameters, incorporating an
EOB or Taylor Hamiltonian with Pad\'e or Taylor expanded flux.  This
method was compared with previous PN methods
\cite{Brugmann:2008zz, Husa:2007rh} and eccentricities comparable to
those in Ref.~\cite{Husa:2007rh} are obtained for equal-mass non-spinning systems.
Additionally, low eccentricities are obtained for systems with
unequal masses and spins for a small selection of spin configurations.

{\em High spins:} Both the Bowen-York and CTS initial data approaches 
have intrinsic limits on the
spins of the black holes which evolve from the initial data.  This issue has been addressed recently in
Lovelace et al.~\cite{Lovelace:2008tw}, where a comprehensive overview of the problems
of high spin initial data is also presented.  Whilst the dimensionless
spins of the black holes on the initial data slice for Bowen-York and CTS
initial data are found to be 0.9837 and 0.99 respectively, this spin quickly
relaxes to about 0.93 after a short period of time (see also
Dain et al.~\cite{Dain:2008ck} where this was previously found for the
Bowen-York case).  
A third type of initial
data is considered, based on superposed Kerr-Schild solutions (with
constraint solution), and this is shown to allow relaxed spins of
above 0.99.  Note that a modification to the standard puncture
approach for head-on collisions can be used to produce high-spin
puncture data as well \cite{Dain:2000hk, Hannam:2006zt}, and for the
single black hole case \cite{Liu:2009zz}, though this
has not yet been studied for the orbiting case.

{\em Constraint solution:} When setting up a configuration of black holes for numerical
simulation with Bowen-York initial data, it is necessary to solve the
Einstein constraint equations, which are elliptic equations on the
initial data slice.  This can be computationally expensive, and the
most commonly used solver code for this, \code{TwoPunctures}
\cite{Ansorg:2004ds}, uses a bi-spherical coordinate system which is
specifically adapted to the case of two black holes.  To increase
flexibility, for example to investigate the possibility of simulating
more than two black holes (see Ref.~\cite{Campanelli:2007ea} for an alternative
approach), in Bode et al.~\cite{Bode:2009fq} an approximate analytic method was
used to solve the constraint equations, based on Ref.~\cite{Faye:2003sw}.
It was found that the main effect of this approximation was to alter
the eccentricity of the resulting evolutions, and when only the merger
and ringdown are in-band, the mismatch between the waveforms with the
exact and approximate constraint solution schemes was $< 10^{-2}$.

{\em Modifications to puncture initial data:} Several studies of the
properties of commonly used initial data have recently been
performed. In Brown \cite{Brown:2009ki}, the propagation of modes in the
puncture geometry is studied with spherically symmetric BSSN
evolutions, and it is found that a potentially problematic numerical
reflection of perturbations in the geometry does not occur due to a
decoupling of the modes of the system.  When initial data is
constructed in the puncture technique, the coordinate conditions used
in the evolution cause the geometry to change during the very first
portion of the evolution and settle to the so-called {\em trumpet}
solution.  In both Hannam et al.~\cite{Hannam:2009ib} and Immerman and
Baumgarte \cite{Immerman:2009ns},
initial data is constructed which is already in the trumpet form.
However, no significant difference is found in Ref.~\cite{Hannam:2009ib}
between the evolution of trumpet and standard puncture data. In
current BBH simulations, the evolution is usually performed on spatial
slices which reach to spatial infinity.  However, gravitational
radiation is properly defined only at future null infinity (see the
discussion in Sec.~\ref{sec:wavemethods}).  An alternative is to
construct slices which asymptote to null infinity rather than spatial
infinity, and one way to do this is to use {\em hyperboloidal} slices.
In Buchman et al.~\cite{Buchman:2009ew}, Bowen-York initial data including unequal
masses, spins and boosts is generalised to such hyperboloidal slices.
The evolution of this initial data should allow extraction of
gravitational radiation at future null infinity, removing a common
source of error in BBH simulations.

\subsection{Boundary Conditions}

Typical NR BBH simulations begin with a three-dimensional initial data
set and evolve this forwards in time.  Due to the difficulties of
simulating an infinite spatial domain with finite numerical resources,
a finite domain is usually simulated in combination with an artificial
boundary condition on the exterior of the domain.
Ideally, these boundary conditions should result in a solution which
is indistinguishable from an evolution with an infinite spatial
domain.  This can only be achieved approximately, and the boundary
conditions should have certain properties to give a useful solution.
Firstly, they should not contaminate the solution with unphysical
gravitational radiation, either due to reflections of the waves
generated by the binary, or due to radiation generated by the boundary
condition itself.  Secondly, they should be {\em constraint
  preserving} to yield a result which is a solution to the Einstein
equations.  Thirdly, they should result in a {\em well-posed initial
  boundary value problem}.  This is a mathematical property which is a
necessary condition for the numerical schemes used to be formally
stable.  These three properties are often only approximately
satisfied, but the effects can be included in estimated error bars on
the physical results of the simulation.

Current evolutions using the \code{SpEC} code in the Generalised Harmonic
formulation such as those in Refs.~\cite{Boyle:2007ft, Scheel:2008rj} use an
outer boundary condition designed to satisfy the Einstein constraint
equations as well as to minimise the incoming 
gravitational radiation \cite{Rinne:2007ui, Lindblom:2005qh,
  Rinne:2008vn}, though these are not designed to be mathematically
well-posed (a necessary condition for formal stability of the problem
under small perturbations).  Recently, in Winicour \cite{Winicour:2009dr},
boundary conditions for the Generalised Harmonic formulation are
presented which are constraint-preserving and of the Sommerfeld type,
which limits reflections of waves off the boundary.  Additionally, the
initial boundary value problem is well-posed.

BSSN simulations generally apply an outgoing radiative boundary
condition by modelling the solution at the boundary as a radially
propagating outgoing wave \cite{Alcubierre:2002kk}.  This approach is
designed to reduce reflections of the outgoing radiation into the
numerical grid which would contaminate the solution.  However, it is
neither mathematically well-posed nor does it respect the Einstein
constraint equations.  It is expected that the errors introduced are
smaller than the errors due to finite differencing, but this has not
been comprehensively studied for long simulations.  One approach is to
place the outer boundary sufficiently far from the binary that the waveforms measured
at finite radius are out of causal contact with the outer boundary,
and this was the approach taken in the early short simulations as
well as in modern simulations for which this is sufficiently inexpensive (for
example those in \cite{Pollney:2009yz}). In N\'u\~nez and Sarbach \cite{Nunez:2009wn}, new
boundary conditions for the BSSN evolution system are proposed which
exactly satisfy the constraint equations, specify approximately zero
incoming radiation, and are stable (the initial boundary value problem
is well-posed) to linear order.  It is hoped that these boundary
conditions can be used in evolutions in future to avoid the
above-mentioned problems.

\subsection{Gravitational Wave Extraction}
\label{sec:wavemethods}

Mathematically, gravitational radiation from a BBH inspiral and merger
is defined at future null infinity; i.e. the part of the spacetime
towards which all null rays, such as light or GWs, propagate. Current
NR BBH codes are based on a Cauchy, 3+1, formalism, which means that
initial data is specified on a (portion of a) 3-dimensional spacelike
slice and then evolved forwards in time.  This leads to a sequence of
spacelike slices on which the solution is known.  GWs must be read off
at a finite spatial radius, and this introduces an error in the
waveform.  An obvious solution is to compute the radiation at several
radii and extrapolate the waveform to infinite radius as a function of
retarded time from the source.  It was shown in Hannam et al.~Ref.~\cite{Hannam:2007ik}
that this is not a trivial procedure, and a detailed analysis and
successful extrapolation of the inspiral portion was presented in
Boyle et al.~Ref.~\cite{Boyle:2007ft}.  A failure to extrapolate waveforms from the \code{SpEC}
code can lead to a phase error of 0.5 radians \cite{Boyle:2007ft}, or
a mismatch of $\sim 10^{-2}$ \cite{Boyle:2009vi}, if using waveforms
extracted at $r = 50 M$ only.

A detailed study of two different techniques for performing this
extrapolation is reported in Boyle and Mrou\'e \cite{Boyle:2009vi}.  The first method
involves extrapolating the amplitude and phase of the complex waveform
at each instant of time using various orders of polynomial
extrapolation.  Essentially, this is the technique which is now common in
the literature, though in Ref.~\cite{Boyle:2009vi} various refinements are
proposed which are shown to improve the quality of the extrapolation.
The second method, based on ideas in Ref.~\cite{Hannam:2007ik},
involves expressing the amplitude and time as a function of the phase
and then performing the extrapolation.  Error estimates are made, and
the two methods are shown to agree within these error estimates.
However, there were problems with the convergence of both methods
after the merger which were not properly understood, and future work
will investigate these.  This study applies only to the waveforms from
the \code{SpEC} code, and the same conclusions might not hold for waveforms
produced by other codes.  Generally, numerical relativists have found
that it is necessary to be extremely careful with the extrapolation
procedure, and, as pointed out in Ref.~\cite{Boyle:2009vi}, there are still
issues to be resolved.

In Pollney et al.~\cite{Pollney:2009ut}, wave extraction techniques at finite radius
are studied and it is shown that the conventional extrapolation of
waveforms extracted at low radii is consistent with extracting at $r =
1000 M$.  The use of such a large extraction radius is possible
because of the constant angular resolution in the NR evolution code,
as demonstrated for lower extraction radius in
Refs.~\cite{Pfeiffer:2007yz,Boyle:2007ft}.  The outer boundary can be placed
out of causal contact with the wave extraction sphere without incurring inordinate computational cost. The peeling
property of the Weyl scalars is also studied, and while $\psi_4$,
$\psi_3$ and $\psi_2$ decay with distance from the source with the
expected exponent, the situation for $\psi_0$ and $\psi_1$ is more
complicated, and different exponents are obtained.  One suggested
explanation for the differing exponents is coordinate and frame
effects in the wave extraction procedure, but it is also pointed out
that the peeling property might not be expected to hold even for
single spinning Bowen-York black hole spacetimes, and by assumption it would
not then hold for binaries either.

 
Instead of performing a traditional Cauchy evolution, where the
solution is evolved along timelike directions, it is possible
to evolve along null directions.  This is called {\em characteristic} evolution.
This cannot be done straightforwardly
in the neighbourhood of
the black holes as it leads to caustics in the solution, but it can be
done in regions sufficiently far from them.  The Cauchy-Characteristic
Extraction (CCE) method combines the two approaches.
Metric data is read at a sufficiently large radius from the
solution generated by a Cauchy code and is used as input data for a characteristic
evolution.  In this way, the waveform can be evolved all the way to
future null infinity, removing ambiguities due to
coordinate conditions or insufficient distance from the source.  This
has been implemented in Reisswig et al.~\cite{Reisswig:2009us,Reisswig:2009rx} and
represents the first unambiguous determination of GWs at future null
infinity for a NR BBH simulation.  Note that the outer boundary
condition
of the Cauchy evolution can still contaminate the waveform, but in
this work the outer boundaries were causally disconnected from the
CCE extraction worldtube, avoiding this problem.
It was found that the waveform
agreed with the extrapolated finite-radius waves within 0.2\% in
amplitude and within 0.01 radians in phase.  It was also determined
that the difference between extrapolated and CCE waveforms was not
significant for gravitational wave detection, but that for Advanced
LIGO, Advanced VIRGO and LISA, these differences might be important
for parameter estimation.

\subsection{Numerical Methods}


For numerical relativity evolution purposes, the four-dimensional BBH
spacetime
is usually foliated with three-dimensional
spacelike slices.  Each of these slices can be split into two regions
with very different computational requirements.  The region around the
black holes has a complicated geometry reflecting the shapes of the
horizons, but the region far from the black holes consists only of
gravitational radiation.  Since the radiation propagates essentially
radially, it requires a constant angular resolution to resolve it.
The majority of BBH NR codes today use Cartesian-type coordinates
everywhere in the grid.  These have the advantage of simplicity, as
only a single coordinate patch is required to cover the entire
simulation domain.  However, they are not efficient,
as they lead to an increasing angular resolution with radius.  If
box-in-box mesh refinement is employed, this inefficiency can be
partially addressed, but only at the expense of significantly reduced
radial resolution.  

In Pazos et al.~\cite{Pazos:2009vb}, multiple locally Cartesian coordinate patches
adapted to the black holes and exterior spacetime were used with finite differencing
methods and Generalised Harmonic evolution equations to compute an
equal mass non-spinning BBH inspiral using excision techniques.  The
scalability and convergence were good, though the simulation developed
instabilities before the merger. In Pollney et al.~\cite{Pollney:2009yz}, a similar
patch structure was used for the exterior spacetime, but adaptive
mesh refinement (AMR) with Cartesian grids was
used for the region surrounding the black holes.  The entire inspiral, merger
and ringdown was simulated.  In both these works, the fact that the
exterior spacetime is resolved using constant angular resolution means that the
computational cost in this region scales linearly with the radius.
This means that the outer boundary of the domain can be placed out of
causal contact with the region of physical interest for only moderate
computational cost.  This bypasses the problem that the correct outer
boundary conditions for the BSSN formulation for the BBH problem
are not known.


While most NR codes use finite differencing methods with global
Cartesian coordinates, the \code{SpEC} code
uses multiple coordinate patches as well as spectral
methods.  For simple equations, these methods can be shown to be
significantly more accurate (exponential rather than polynomial
convergence) for a given computational cost, and indeed, the quoted
accuracy and efficiency on the waveforms from the \code{SpEC} code is
impressive.  In Szil{\'a}gyi et al.~\cite{Szilagyi:2009qz}, a new gauge condition is
presented along with a new grid structure (touching grids with penalty
boundary conditions, rather than the previously used overlapping
grids), as well as a new method of dynamically adapting the grids to
the shapes of the black holes.  With these new techniques, \code{SpEC} is able to
handle a variety of configurations without fine-tuning the simulation
parameters on a case-by-case basis during the merger phase as was
previously required.  Simulations of mass ratios of $q = 2$ and
dimensionless spins of 0.4 are presented.

\section{Challenges and Conclusions}
\label{sec:challengesandconclusions}

In this work, I have given a brief overview of the current state of
BBH simulations in NR.  It is now possible to simulate BBH systems
over periods as long as 15 orbits for low mass ratios and moderate
spins, or to simulate a smaller number of orbits for mass ratios up to
1:10 or dimensionless spins up to $0.92$.  Pushing these simulations
to higher numbers of orbits, mass ratios and spins can be achieved but
at significant computational cost.  For arbitrary spin configurations,
it is not known how many black hole orbits are needed from NR before a
sufficiently accurate complete waveform using input from analytic
methods can be constructed.  In order to span the parameter space of
initial spins and mass ratios, many simulations will be needed.  These
are still very expensive to perform, taking weeks or months on up to a
thousand CPUs for high mass ratios and large numbers of orbits.

Approximately speaking, the computational cost of a BBH simulation
scales with the mass ratio $q$, based on the resolution requirement of
the smaller black hole alone.  For an accurate equal mass, non-spinning 8
orbit simulation which takes on the order of 200 000 CPU hours, a
simple estimate shows that scaling this to a mass ratio of $q = 10$
would take 2 million CPU hours, and this is probably an
underestimate, as the fact that the waveform gets longer with
increasing $q$ has not been considered, nor the fact that the higher
resolution required around the smaller hole will not always come with
perfect computational scaling.  Since many simulations for different
spin configurations would be required to construct or tune template banks used in
gravitational wave data analysis for detection and parameter
estimation, this is clearly infeasible (computer resources for NR BBH
groups for an entire year are usually of the order of a few million
CPU hours).

There are also theoretical issues which need to be better understood.
As explored in Ref.~\cite{Boyle:2009vi}, the extrapolation of finite radius
waveforms is not yet perfect, especially in the neighbourhood of the
merger.  Cauchy-Characteristic Extraction for BBH systems
\cite{Reisswig:2009us,Reisswig:2009rx} should in principle solve this
problem, but its use is not yet widespread.  It is also not yet clear
exactly how to construct accurate waveforms from PN and NR
results when the PN waveforms come with their own intrinsic errors.
It has been shown \cite{Buonanno:2009zt} that most of the
different PN approximants can be used equally well for detection with
ground-based detectors of systems below a certain total mass (12
$M_\odot$), indicating that the errors in each of these approximations are
not significant for the low frequencies for which such systems are
in-band. However, for higher masses, and for parameter estimation, the
PN errors become significant, and assuming that they are zero when
constructing PN-NR waveforms is not necessarily the best
choice.

The generation of the gravitational recoil around the BBH merger is
not understood in precise terms.  Simple analytic models have been
tested \cite{Campanelli:2007ew, Baker:2008md, Lousto:2008dn,
  Lousto:2009mf} and shown to fit well with numerical data.  However,
the models are {\em a posteriori} empirical fits to forms
qualitatively inspired by PN predictions in a region where the
agreement with PN is poor.  Also, the crucial identification of the
orbital plane is gauge dependent. To get a robust, coordinate
independent understanding of this recoil will require detailed
understanding of the physics in the region near the black holes.  The
work of Lovelace et al.~\cite{Lovelace:2009dg}, where the linear momentum in this
region is computed, may be a starting point for this.  Previous work
\cite{Schnittman:2007ij} has also analysed the features of the recoil,
showing that a Newtonian model containing some input from the
simulations combined with a Kerr quasinormal mode expansion
is able to reproduce the numerical results.
There has also
been recent work \cite{Tiec:2009yf,Tiec:2009yg} combining PN with
perturbation theory in the close-limit approximation
\cite{Price:1994pm} where the recoil for non-spinning black holes is computed
and agrees moderately well with NR results.  The extension of this
work to black holes with spins should lead to greater insight into the
generation of the recoil.

There have been recent theoretical advances which are not yet
routinely incorporated in numerical simulations.  More astrophysically
realistic initial data prescriptions have been given
\cite{Kelly:2007uc, JohnsonMcDaniel:2009dq}, and initial
evolutions performed \cite{Kelly:2009js}.  It is not clear to what
extent the constraint violations present in these initial data sets
will impact their usefulness, and this needs to be studied further.  A
method for constructing initial data for almost maximally spinning black holes
has also been proposed \cite{Lovelace:2008tw} which goes beyond the
capabilities of the Bowen-York initial data currently used by many
groups.  Two possibilities for improving the physical accuracy of
gravitational waveforms from BBH simulations have recently been explored:
hyperboloidal slicing conditions \cite{Buchman:2009ew},
in which future null infinity is approached in the computational
domain, and Cauchy Characteristic Extraction
\cite{Reisswig:2009us,Reisswig:2009rx}, where the waveforms
are evolved to future null infinity using 
a separate null code.  Whilst it is shown that the
CCE waveforms are consistent with currently-used extrapolation techniques,
this picture may change as simulations become more accurate.

In summary, there are many interesting problems to be addressed in
vacuum NR.  In terms of physics, these include gaining improved
understanding of the strong field geometry around the black holes, and how
this affects the final black hole produced in the merger.  Computationally,
increased efficiency for long inspirals, very high spins, and higher
mass ratios are the main challenges.  Finally, a strong interaction
with the community of gravitational wave data analysts will be
necessary to fully leverage the exciting progress that has been made
in NR to date.

\begin{acknowledgments}
  It is a pleasure to thank M.~Hannam, F.~Herrmann, H.~Pfeiffer,
  E.~Schnetter and L.~Rezzolla for reading a draft of this manuscript
  and providing detailed and helpful comments and criticism.  All
  errors and omissions remain the responsibility of the author.
\end{acknowledgments}

\vfill

\bibliography{references2}

\end{document}